\documentclass[twocolumn,prl,showpacs,floatfix]{revtex4}

\usepackage{graphicx}
\usepackage{epsfig}
\usepackage{rotating}
\usepackage{amsmath}
\usepackage{amsfonts}
\usepackage{amsfonts}
\usepackage{amssymb}
\usepackage{enumerate}
\usepackage{longtable}
\usepackage{dcolumn}
\usepackage{bm}

\begin{document}
\title{Phase transitions in spin-orbital models with spin-space anisotropies for iron-pnictides:
A study through Monte Carlo simulations}

\author{Ryan Applegate}
\affiliation{University of California Davis, CA 95616, USA}
\author{Rajiv R. P. Singh}
\affiliation{University of California Davis, CA 95616, USA}
\author{Cheng-Chien Chen}
\affiliation{Stanford Institute for Materials and Energy Science, SLAC National Accelerator Laboratory, Menlo Park, California 94025, USA}
\address{Geballe Laboratory for Advanced Materials, Stanford University, Stanford, California 94305, USA}
\author{Thomas P. Devereaux}
\affiliation{Stanford Institute for Materials and Energy Science, SLAC National Accelerator Laboratory, Menlo Park, California 94025, USA}
\address{Geballe Laboratory for Advanced Materials, Stanford University, Stanford, California 94305, USA}


\date{\today}

\begin{abstract}
The common phase diagrams of superconducting iron pnictides show interesting material specificities 
in the structural and magnetic phase transitions. In some cases the two transitions are separate and second order, while in others 
they appear to happen concomitantly as a single first order transition.
We explore these differences using Monte Carlo simulations of a two-dimensional Hamiltonian
with coupled Heisenberg-spin and Ising-orbital degrees of freedom.
In this spin-orbital model, the finite-temperature orbital-ordering transition results in a tetragonal-to-orthorhombic
symmetry reduction and is associated with the structural transition in the iron-pnictide materials.
With a zero or very small spin space anisotropy, 
the magnetic transition separates from the orbital one in temperature, and the orbital transition is found to be in the Ising universality class.
With increasing anisotropy, the two transitions rapidly merge together and tend to become weakly first order.
We also study the case of a single-ion anisotropy and propose that the preferred spin-orientation along the antiferromagnetic direction in these materials is driven by orbital order.
\end{abstract}

\pacs{74.70.Xa,75.25.Dk,75.40.Cx,75.40.Mg}

\maketitle

\section{Introduction}
Several parent phases of the superconducting iron-pnictide materials show an interesting interplay of structural, magnetic
and orbital degrees of freedom. \cite{discovery_1111, discovery_122, Review_Johnston} 
These quasi two-dimensional (2D) materials share a similar phase diagram, where
a tetragonal paramagnetic phase at high temperatures transitions into an orthorhombic, antiferromagnetic
phase at low temperatures.\cite{phase_1111_Zhao, phase_1111_Rotundu, phase_1111_secondorder, phase_122_Chen, phase_122_Chu, phase_122_Nandi} 
The square-lattice of iron atoms develop magnetic order at 
wavevector ($\pi,0$), which corresponds to antiferromagnetic (AFM) alignment of spins
along one of the nearest-neighbor direction (x) and ferromagnetic (FM) alignment 
along the other (y).\cite{moment_1111_Cruz, moment_1111_Chen, moment_1111_Zhao, moment_Ba122_Huang, moment_Ba122_Kofu, moment_Ba122_Su, moment_Ba122_Goldman, moment_Sr122_Kaneko, moment_Sr122_Zhao, Review_magnetism}
The orientation of the ordered spin moments is tied to antiferromagnetism and 
points along the AFM direction.\cite{moment_Ba122_Huang, moment_Ba122_Kofu, moment_Ba122_Su, moment_Ba122_Goldman, moment_Sr122_Kaneko, moment_Sr122_Zhao, Review_magnetism}

While lattice distortions 
are typically quite small in iron pnictides, the observed spin-wave spectra from neutron scattering
suggest a robust, possibly sign changing anisotropy in the exchange constants 
along the x and y directions.\cite{neutron_anisotropy_Ca122, neutron_anisotropy_Ba122, neutron_anisotropy_Sr122} 
Various transport, optical, and spectroscopic measurements also show substantial emergent anisotropies in the 2D xy plane. \cite{transport_Chu_magnetic, transport_Chu_mechanical, transport_Tanatar, transport_Ying, transport_Kuo,
optical_Dusza, optical_Lucarelli, optical_Nakajima, STM_Chuang, Ian_review} 
In particular, an orbital polarization associated 
with the occupation of $d_{xz}$ and $d_{yz}$ orbitals has been observed.\cite{ARPES_Shimojima, ARPES_Lee, ARPES_Wang, ARPES_Yi}
These anisotropies in some cases can persist up to high temperatures and have been identified with long 
and sometimes short range Ising-nematic order.\cite{SiAbrahams,Kivelson} 

Despite the above similarities,  there are also substantial material-specific differences. 
The parent compounds of the 1111 family (RFeAsO, with R a rare earth element) of iron pnictides undergo two separate second order phase transitions, where the structural transition is followed by a magnetic transition at a lower temperature.\cite{phase_1111_secondorder} On the other hand, in the 122 family (AFe$_2$As$_2$, with A an alkaline earth element) the two transitions appear to occur at the same temperature.\cite{phase_122_Chen, phase_122_Chu, phase_122_Nandi}

More recent measurements revealed that in the undoped BaFe$_2$As$_2$, the structural and magnetic transitions are slightly separated by less than one Kelvin.\cite{Birgeneau_Ba122, Kim_Ba122, Birgeneau_review} In that case, the structural transition starts as second order, and at a slightly lower temperature there is a first order jump in the lattice distortion with a concomitant first order magnetic transition. This feature is not generic to all the 122 family of iron pnictides. In particular, there is strong evidence showing a largely first order phase transition in CaFe$_2$As$_2$ and SrFe$_2$As$_2$, where the structural and magnetic phase transitions coincide.\cite{Firstorder_Ca122,Firstorder_Sr122, Suchitra_Sr122}

There have been many proposals for the mechanism driving these transitions.
These include, (i) emergent Ising nematic orders in
frustrated spin systems,\cite{Yildirim, SiAbrahams, Kivelson, Sachdev,
Kotliar, MazinJohannes, zpYin, Han, Antropov, uhrig, Kamiya} (ii) orbital order \cite{Kruger, Singh, Chen, Phillips, WeiKu, OO_Thalmeier, OO_Yanagi, OO_Bascones, OO_Kontani,nevidomskyy} (iii) coupling to lattice 
degrees of freedom,\cite{Schmalian,paul} and (iv) symmetry breaking associated with
fermi-surface effects in an itinerant system.\cite{paul, Tesanovic,  eremin, Brydon, Pomerancuk_Zhai, Fernandes_new} On symmetry grounds one cannot
distinguish between different pictures, since the different degrees of freedom
lead to same broken symmetries and they are all present to some extent
and coupled to each other. Thus, detailed quantitative studies are
important to establish the role played by different mechanisms. 

In this paper, we wish to study the scenario where orbital order is
the primary driving mechanism for the finite temperature transitions.
We investigate the properties of 
a  spin-orbital model, where the spin and orbital degrees of freedom are coupled by a Kugel-Khomskii like
mechanism.\cite{kugel-khomskii} In the model, the local orbital occupation modulates the spin exchange
constants.
Once the orbitals are ordered,  collinear antiferromagnetism can develop and 
anisotropic exchange constants in the x and y directions $J_{1x}\neq J_{1y}$ result. 
However, we note that a model containing only effective Heisenberg spin interactions 
($\sum_{ij}J_{ij} \vec{S}_i \cdot  \vec{S}_j$)
is still \emph{isotropic in spin space},
since the energy of the system does not depend on the direction of magnetization with
respect to the crystal axes.

A simple mean-field treatment of our spin-orbital model
suggests that the spin and orbital orderings occur simultaneously as a single phase transition,
which can be first or second order depending on the exchange couplings. 
However, such a treatment neglects long-wavelength fluctuations which can drive the spin-ordering temperature to zero.
To study the effects of fluctuations, we employ large scale Monte Carlo simulations
by treating the spin and orbital variables classically, which should be sufficient for finite-temperature phase transitions.
The Monte Carlo results indicate that if spin rotational invariance is preserved,
at finite temperatures there is only one orbital ordering transition which belongs to
the 2D Ising universality class. In this case, long-range spin order only occurs at $T=0$, 
in accord with the Mermin-Wagner theorem. 
However, a small spin space anisotropy ($~5\%$) will
bring the magnetic transition temperature up to the orbital one.
With increasing anisotropy the
coupled spin-orbital transition tends to become first order.
These results are reminiscent of the observed behaviors of different families of iron pnictides.

Our study neglects three-dimensional (3D) couplings, studied for example in Refs. \onlinecite{Kivelson}, \onlinecite{Antropov} and \onlinecite{Kamiya}.
3D couplings have a similar effect as spin space anisotropies in that they both can result in a 
finite-temperature magnetic transition.
However, in general they will lead to different universality classes for the transitions.
While 3D couplings could be more important in some materials (for example
within the 122 family), spin space anisotropy may be more important in others.
In some materials
the magnetization has been reported to obey 2D Ising universality behavior.\cite{wilson10}
Even if the ultimate transition is weakly first order in these materials, the reported fluctuations appear more 2D. 
This provides a motivation for our choice of anisotropy over 3D couplings.

Studying spin space anisotropy also allows us to address the orientation of the ordered moments.
In a quasi-two dimensional material, one would expect the uniaxial anisotropy to point out of the plane
and the spins should be equally likely to point along any direction in the plane. However, this can
change with orbital polarization.
In transition metal compounds, ligand crystal-field splitting can lift the degeneracy of the transition metal $3d$ orbitals. In this case, the orbital moments are usually quenched and there may be no preferred spin directions.
However, relativistic spin-orbit coupling can induce a non-zero orbital angular momentum, which
accompanied by an orbital polarization (such as a preferential occupation of $d_{xz}$ over $d_{yz}$ orbitals) 
can lead to a single-ion anisotropy term and an anisotropic $g$-factor in the xy plane.
In this case, excess population of $d_{xz}$ orbitals can favor spins pointing along
the x-axis, while excess population of $d_{yz}$ orbitals will favor spins
pointing along the y direction.
We propose that in iron pnictides the single-ion anisotropy term in the xy plane
is related to orbital order and since it is also tied to AFM it leads to spins
pointing along the AFM direction.

The outline of the paper is as follows. In section II we introduce our
model and develop a simple mean field theory. In section III the Monte Carlo
results are discussed.
In section IV we discuss the implication of our study for the
iron pnictide materials and in section V we summarize our work.
Details of the Monte Carlo method are presented in the appendix.

\section{II. Spin-orbital Model}
The spin-orbital model is given by the Hamiltonian,
\begin{align}\label{eqn:spin_orbital}
        H=&\sum_{i} (J_1 n_i n_{i+\hat{x}}-J_F)\vec{S_i}\cdot \vec{S}_{i+\hat{x}} \\ \nonumber
		  +&\sum_{i} (J_1 (1-n_i) (1-n_{i+\hat{y}})-J_F) \vec{S_i}\cdot \vec{S}_{i+\hat{y}} \\ \nonumber
		  +&\sum_{<<i,j>>} J_2 \vec{S_i}\cdot \vec{S_j}~.
\end{align}
Here $\vec{S}_i$ are classical Heisenberg spins on a square-lattice, 
$n_i$ are classical Ising variables 
that take values $0$ or $1$, and $<<i,j>>$ signifies summing on next nearest 
neighbor pairs of the square-lattice.
In a classical system the spin magnitude $S$ is not important in determining the phase transitions,
and for our discussion we set $S=1$.
Physically, the variables $n_i$ represent
the preferential occupation of $d_{xz}$ ($n_i=1$) or $d_{yz}$ 
($n_i=0$) orbitals.
The model has tetragonal symmetry. However, $n_i=1$ 
(occupation of $d_{xz}$ orbitals) favors AFM 
order along the x-axis, whereas $n_i=0$ (occupation of $d_{yz}$ orbitals)
favors AFM order along
the y-axis. We have added an orbital independent FM nearest neighbor
interaction $J_F$ and used $J_2=0.4J_1$ and $J_F=1/6 J_1$.
This set of parameters corresponds to the neutron scattering
observation that the spin-wave spectra is better fit with an AFM
exchange along one direction and a weak FM exchange along the 
other.\cite{Applegate} 
The latter could arise from double exchange\cite{Phillips} or from the orbital
geometries.\cite{Singh} But, its sign or magnitude is not crucial for the 
phase transitions we report here. Spin-space anisotropies will be
introduced later when we discuss the Monte Carlo simulations.

The ground state of this model breaks tetragonal symmetry. It has a ferro-orbital 
order, all $n_i = 1$ or $n_i = 0$, corresponding to nearest neighbor exchanges which are AFM 
along one axis and FM along the other. The ground state has $(\pi,0)$ spin 
order when $n_i=1$ and $(0,\pi)$ spin order when $n_i=0$.

We note here that in iron pnictides, the low temperature orbital polarization is found to be incomplete,
where the occupation number is not strictly one or zero.\cite{ARPES_Shimojima, ARPES_Lee, ARPES_Wang, ARPES_Yi}
A partial orbital polarization can result from the itinerant electron degrees of freedom, or from quantum fluctuations in the orbital variables.
The role of orbital order in driving the structural and magnetic transitions of iron pnictides indeed has been discussed based on an itinerant electron perspective 
using multi-orbital Hubbard Hamiltonians.\cite{OO_Thalmeier, OO_Yanagi, OO_Bascones, OO_Kontani}
Our approach of studying a Kugel-Khomskii type spin-orbital model can be viewed as the strong  coupling limit of such Hamiltonians.
While we leave out the charge degrees of freedom which are important in describing for example transport properties, our model should still capture the key physics of magnetism and finite temperature phase transitions.

Below we first develop a mean-field theory for the phase transitions of the spin-orbital model under consideration.
We  set $n_i=(1+\sigma_i)/2$ and assume a mean-field
Hamiltonian of the form 

\begin{equation}
{\cal H}_{MF} = -\sum_{i1} B_1^{i1} S_{i1} - \sum_{i2} B_2^{i2} S_{i2} 
-h\sum_i \sigma_i 
\end{equation}
where the first sum runs over sublattice one, the second over sublattice two,
and the third over all the spins in the lattice.
$B_1$ and $B_2$ are the staggered fields on the two sublattices
and $h$ is a field that couples to orbital order. Focusing on the ($\pi,0$)
order, we let $m=\langle S_i\rangle$ and $n=\langle\sigma_i\rangle >0$. We find
for $i=1,2$
\begin{equation}
B_i=2m(J_1 n +2J_2), \qquad h=J_1m^2,
\end{equation}
leading to the mean-field equations,
\begin{equation}
m=L(2\beta m (J_1 n + 2 J_2)),
\end{equation}
and
\begin{equation}
n=\tanh{(\beta J_1 m^2)}.
\end{equation}
Here, $L(x)$ is the Langevin function $\coth{x}-1/x$.
These equations lead to a simultaneous transition and an
orbital ordered AFM phase. It is a second order phase
transition, with a transition temperature of $4 J_2/3$, provided
$J_2>$ constant $J_1$. The transition becomes first order when
$J_1$ exceeds $J_2$ (the case of interest in the pnictides).

While mean-field theory can not be quantitatively valid because
of the divergent infrared fluctuations in the spin variable, which
push the spin ordering transition temperature to zero, we will see
that the mean-field results correctly capture the following physics:

1. Non-zero magnetic order produces an ordering field for the orbital
degrees of freedom. Hence, whenever there is magnetic order present,
orbitals symmetry will also be broken. Thus, orbital transition can not
happen below the magnetic ordering transition.

2. Without some order of the magnetic degrees of freedom, the orbitals 
do not interact. Actually, orbital couplings depend on short-range magnetic 
order not long-range magnetic order. This is not allowed for in the mean-field 
theory but will become clear from our later discussion of the Monte Carlo simulations.
Thus, the two transitions are always going to be close in temperature,
unless the magnetic transition is pushed significantly below the
mean-field transition temperature due to additional fluctuations.

3. The orbital ordering temperature is not significantly depressed by the
fluctuations of the spins and our mean field theory provides a
fairly good prediction of the transition temperature. 

4. We will see in the Monte Carlo simulations that the main
role of the long-wavelength spin fluctuations is to
decouple the spin and orbital transitions. The spin transition
temperature is pushed to zero in the absence of spin space anisotropy.
In this case, the orbital transition becomes Ising like and
second order.

5. With significant anisotropy, both the spin and orbital transition 
temperatures rapidly approach the mean-field values, and the transition
has a tendency to become first order for $J_1>J_2$.


\section{III. Results of Monte Carlo Simulations}
In this section, we present the results of Monte Carlo 
simulations with and without spin-space anisotropies.
The details of the Monte Carlo methods as well
as the quantities measured and the expected scaling behavior
are discussed in the appendix.

\subsection{Isotropic Heisenberg Spins}
\begin{figure}[t!]
\includegraphics[width=0.8\columnwidth,clip,angle=-90]{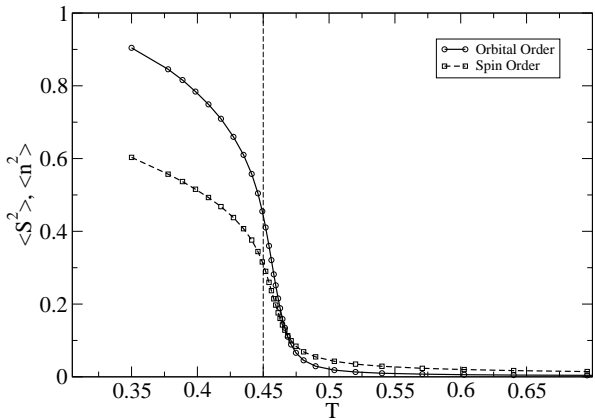}
\caption{Squares of spin and orbital order parameters as a function of temperature
for the isotropic spin-orbital model on a $20\times 20$ lattice.
The vertical line shows the transition temperature $T_c$ (measured in units  of $J_1$), where the
orbitals develop long-range order.}
\label{fig:order_parameters}
\end{figure}

First, we consider the case of isotropic Heisenberg spins.
The squares of spin and orbital order parameters obtained from the
simulation are shown in Fig. \ref{fig:order_parameters}.
We know on general grounds 
that in a 2D system spin rotational symmetry can not be
spontaneously broken at any finite temperature.
However, this is not evident from the plot. The exponential growth
of the spin-spin correlation length  rapidly exceeds the size of the 
system and this creates the impression of long-range order
at a finite temperature. One needs to carefully study the
size dependence. The Binder 
ratios, defined in the appendix, prove useful for this purpose.

\begin{figure}[t!]
\includegraphics[width=0.8\columnwidth,clip,angle=-90]{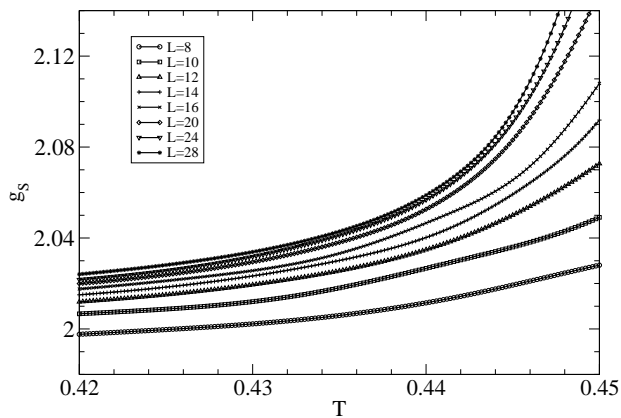}
\caption{
Spin Binder ratio as a function of temperature (measured in units of $J_1$) for different $L \times L$ lattices.
For the isotropic spin-orbital model there are no crossings in $g_S$ at any
temperatures of our simulation. This is consistent with the theory that in a 2D system there is no long range spin order at finite temperatures.}
\label{fig:spin_isotropic_binder}
\end{figure}

\begin{figure}[t!]
\includegraphics[width=0.8\columnwidth,clip,angle=-90]{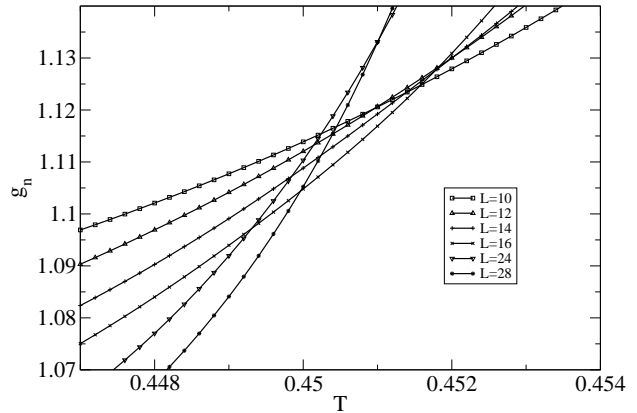}
\caption{
Orbital Binder ratio as a function of temperature (measured in units of $J_1$) for different $L \times L$ lattices.
In contrast to the spin variables, the discrete orbital variables undergo a phase transition, developing long range order at finite temperatures.}
\label{fig:orb_isotropic_binder}
\end{figure}

Figure \ref{fig:spin_isotropic_binder} gives the spin binder ratios, $g_S$, which show no crossings with system size
down to the lowest measured temperature, signifying absence of long range 
order at finite temperatures, in agreement with the Mermin-Wagner theorem.
In contrast the orbital binder ratios $g_n$, shown in Fig. \ref{fig:orb_isotropic_binder},
have clear crossings at 
finite temperatures and we can extract $T_c$ by comparing different system 
sizes. We obtain $T_c/J1=0.450 \pm 0.001$ for the isotropic spin-orbital model.

\begin{figure}[t!]
\includegraphics[width=0.8\columnwidth,clip,angle=-90]{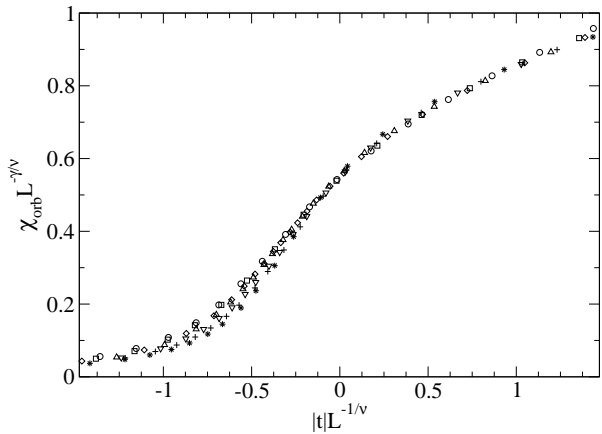}
\caption{The scaling of the universal $\tilde\chi$ versus reduced temperature
$|t|=|(T-T_c)/T_c|$ shows that critical exponents are consistent with the 2D Ising Model.}
\label{fig:critical_orb1}
\end{figure}

\begin{figure}[t!]
\includegraphics[width=0.8\columnwidth,clip,angle=-90]{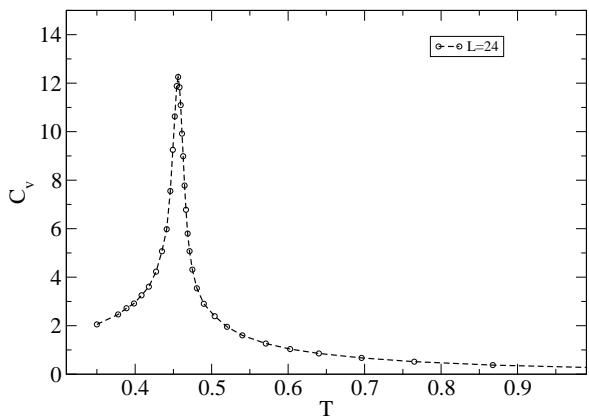}
\caption{Specific heat for the isotropic spin-orbital model under study. 
The sharp peak is consistent with a logarithmic divergence at $T_c$.}
\label{fig:specific_heat}
\end{figure}

In Fig. \ref{fig:critical_orb1} we show the scaling plot for the orbital susceptibility. The
data collapse leads to estimates of critical exponents $\nu=1.01\pm.01$ and
$\gamma=1.75\pm.02$. These exponents are consistent with the 2D
Ising universality class. Figure \ref{fig:specific_heat} shows a plot of the 
specific heat, which grows rapidly near $T_c$. It
is consistent with a logarithmic divergence but with an amplitude significantly larger than that in the pure 2D Ising model.
The amplitude of the specific heat is not universal but is comparable
for the Ising model on different 2D lattices.\cite{houtappel} It is
considerably larger in our model, presumably as the Ising nematic variables
associated with the spins couple directly to the orbitals and enhance 
the amplitude.
We last note that the sharp peak in our specific heat clearly indicates a phase transition and its transition temperature $T_c$.
Therefore, the Monte Carlo simulations of of our spin-orbital are less affected by finite-size effect compared to that 
in the frustrated square lattice $J_1$-$J_2$ model.~\cite{Weber}

\subsection{ Exchange and single-ion anisotropy }
We next consider models with spin space anisotropy by generalizing the scalar product
\begin{equation}
\label{eqn:anisotropy_term}
\vec{S_i}\cdot \vec{S_j} = S^z_iS^z_j + \lambda [S^x_iS^x_j+S^y_iS^y_j]~.
\end{equation}

\begin{figure}[t!]
\includegraphics[width=0.8\columnwidth,clip,angle=-90]{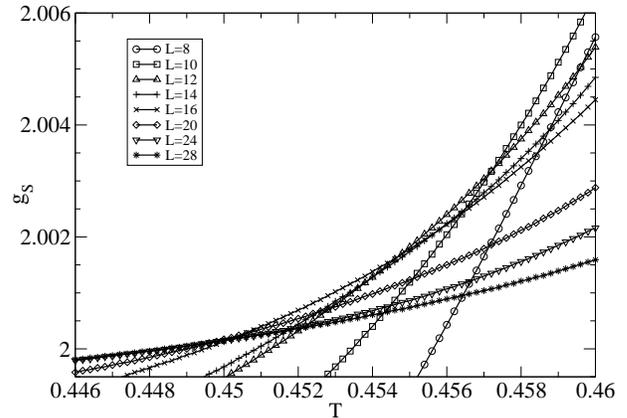}
\caption{
Spin Binder ratio as a function of temperature (measured in units of $J_1$) for different $L \times L$ lattices.
For anisotropic spins with either Ising or single ion anisotropy, finite spin ordering is observed besides orbital order.}
\label{fig:spin_anisotropic_binder}
\end{figure}

\noindent
We study the system for several values of the Ising anisotropy parameter $\lambda$.
As long as $\lambda < 1$, there is only Ising symmetry for the spins, and 
both spins and orbitals can order at finite temperatures.
The effect of the anisotropy on the orbital order is small, and the transition
temperature is raised gradually as $\lambda$ is reduced. In contrast, one can
see a dramatic difference in the Binder ratios for the spin variables.
Comparing to isotropic spins, viz. $\lambda=1$, in Fig. \ref{fig:spin_isotropic_binder}, there are clear
crossings in Fig. \ref{fig:spin_anisotropic_binder}. We can extract $T_c$ for both order parameters using the Binder ratio.

\begin{figure}[t!]
\includegraphics[width=0.8\columnwidth,clip,angle=-90]{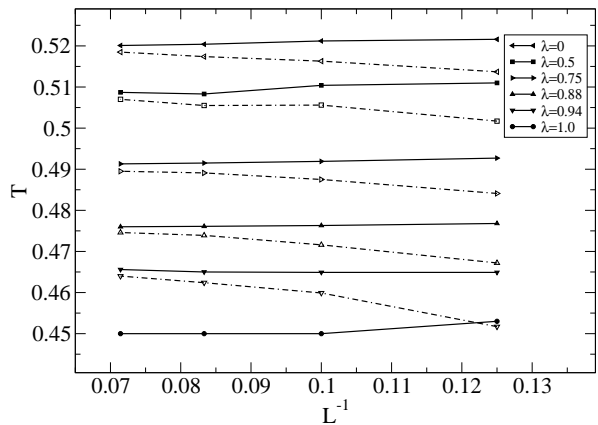}
\caption{From Binder cumulant ratios systematic crossings are located 
for systems of size $L$ and $2L$. These are plotted versus inverse 
system length and extrapolated to
get the thermodynamic $T_c$ (measured in units of $J_1$). In all cases, the orbital crossings (solid) approach 
$T_c$ from above and the
spin crossings (dashed) approach it from below. For $\lambda=1$, spins order only at zero temperature.}
\label{fig:anisotropy_regress}
\end{figure}

We summarize the extraction of $T_c$ over a range of $\lambda$ in Fig. \ref{fig:anisotropy_regress}. 
When $\lambda$ is near (but not equal to) 1, we are in a regime where the spin transition temperature is non-zero but still separated from the orbital transition.
However, when the anisotropy is small (in our case, $|1-\lambda| < 0.1$), computationally it is difficult to distinguish the two transitions.
This shows that the spin transition 
temperature grows very rapidly with increasing anisotropy and it rapidly
merges with the orbital transition. We note that with anisotropy the
transition temperatures are within $10\%$ of the
mean-field value of $0.53J_1$.

\begin{figure}[t!]
\includegraphics[width=1.0\columnwidth,clip,angle=0]{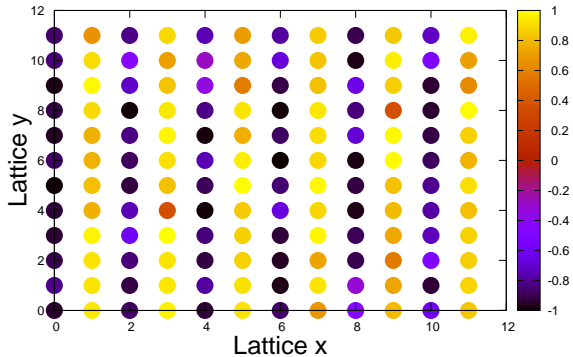}
\caption{In the orbital configuration $\{n_i=1\}$, the spin-orbital model with single ion anisotropy has an AFM exchange along the x direction and favors the spin order collinear with this exchange. This is shown by plotting the $S^x$ component from a typical spin configuration when in the $\{n_i=1\}$ phase. The false color plot represents magnitude of the spin component along the x direction.}
\label{fig:config_single_ion}
\end{figure}

We now introduce a single ion anisotropy, which is tied to orbital order.
\begin{equation}\label{eqn:single_ion_term}
H_{ion} = - D \displaystyle\sum_{i}^N (n_i{S^x_i}^2 + (1-n_i){S^y_i}^2)~.
\end{equation}
In transition metal compounds, ligand crystal-field splitting lifts the degeneracy of the transition metal $3d$ orbitals, and the orbital angular moments are usually quenched. In this case, treating the relativistic spin-orbit coupling as a perturbation to the second order will result in a single-ion anisotropy term that reflects the underlying symmetry of the crystal. Therefore, an orthorhombic structural distortion or a net orbital polarization can lead to a single-ion anisotropy term closely tied to orbital order.
This single ion anisotropy favors spin orientations along the AFM direction.
One ordered configuration observed
in the simulation is shown in Fig. \ref{fig:config_single_ion}.

Similar to the case of Ising-anisotropy (Eq. (\ref{eqn:anisotropy_term})), we have simulated these
systems with different $D$ values. For $ |D| > 0.1$, 
once again we see no separation of the two ordering transitions for spins 
and orbitals. This shows that, like the Ising anisotropy, the uniaxial
anisotropy causes a rapid increase in the spin ordering temperature
and it soon merges with the orbital order.

\begin{figure}[t!]
\includegraphics[width=0.8\columnwidth,clip,angle=-90]{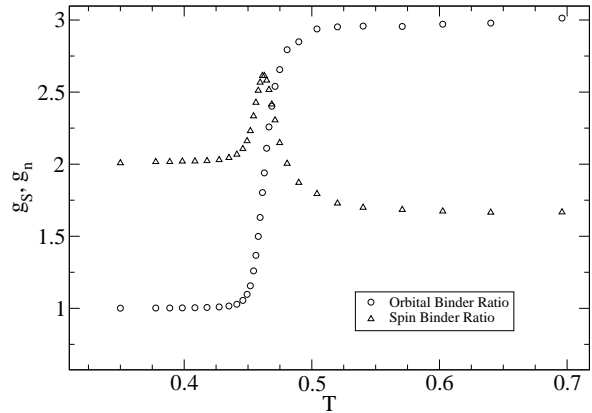}
\caption{For an isotropic ( $\lambda = 1$ ) $24 \times 24$ system, we plot the orbital and spin binder ratios, $g_n$ and $g_S$. $g_n$ behaves smoothly between its limiting values while $g_S$ develops what seems like a divergence at the transition temperature. The development of such an incipient divergence is an indicator of a first order transition.}
\label{fig:binder_divergence}
\end{figure}

In Fig. \ref{fig:binder_divergence}, we show the binder ratios for orbitals and spins. 
Binder ratios for orbitals remain well behaved regardless of the anisotropy introduced in the models. However, there is a clear
incipient divergence in $g_S$, which
is indicative of a first order transition.
In general, we find that the spins have a greater tendency
for a first order transition than the orbitals. The implications of
these results for the pnictides are discussed in the next section.




\section{V. Discussion and Relevance to the Iron Pnictides}

In this section, we use the results of Monte Carlo simulations,
mean field theory and general arguments about 
quasi-2D spin systems
to develop an overall phase-diagram
for coupled spin-orbital systems.
We will then explore the applicability of the phase diagram to the iron-pnictide materials.
The key issues of interest to us are whether there is a single transition
or two separate transitions, and whether each of the transitions is first or
second order.

\subsection{The phase diagram of the spin-orbital model}

The mean-field theory gives a simultaneous spin and orbital transition,
which could be first or second order depending on the exchange couplings. 
Monte Carlo simulations show that
the spin and orbital transitions are practically simultaneous unless
the spin space anisotropy is very small. In the latter case, divergent
long-wavelength fluctuations push the spin transition temperature to zero,
whereas the orbital transition is not significantly affected by these
fluctuations. The transition temperatures observed in the simulations
are within a few percent of the mean-field value of $0.53J_1$ when the
anisotropy is large. As the anisotropy goes to zero, the orbital
transition is reduced by less than $20\%$, whereas the spin
transition is reduced all the way to zero. Even a $5\%$
anisotropy causes a near simultaneous transition.

On general grounds, one knows that in place of long-range order
the correlation length in a 2D Heisenberg spin system stays finite
but grows exponentially as
$\exp{C/T}$ as the temperature is lowered. This implies that if
below some energy scale $\epsilon_0$ these divergent fluctuations are cut off
(due to for example spin space anisotropy or 3D coupling), it will lead to
long-range order and the transition temperature will depend
on the energy scale as $1/\ln{(\epsilon_0/\epsilon)}$,
rising very steeply with increasing $\epsilon$.\cite{CHN-PRB39-2344-1989} 
Our Monte Carlo simulations show that unless the spin transition is significantly suppressed by
fluctuations the spin and orbital transitions would happen together.

The simulation results also show that the isolated orbital transition
is in the universality class of the 2D Ising model. While we have not
been able to observe the isolated finite temperature spin transition
when it is separated from the orbital transition, on general grounds,
we expect it also to be a continuous transition in the universality class
of the 2D Ising model (due to a small but non-zero spin space anisotropy). If the
3D couplings are more important than spin space anisotropy, then the
transition could have a significant crossover region 
in the universality class of the 3D Heisenberg model (but will 
ultimately be in the 3D Ising universality class if a uniaxial spin anisotropy is also present).
When the two transitions come together, the simulations find that
the transitions tend to become first order.

\begin{figure}[t!]
\includegraphics[width=0.8\columnwidth,clip,angle=-90]{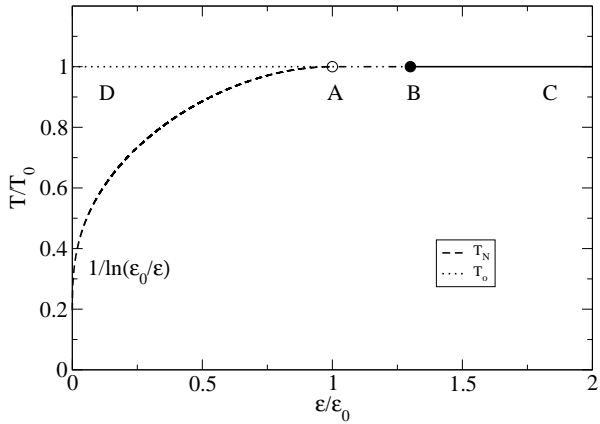}
\caption{Phenomenological phase diagram for the spin-orbital model. 
The exchange energy scale in the problem sets the transition temperature $T_O$ for orbital order,
which in turn drives the structural transition.
$\epsilon_0$ is the energy scale below which long wavelength fluctuations are suppressed.
There are two separate continuous orbital and magnetic transitions for $\epsilon \ll \epsilon_0$ (shown as dotted and dashd lines respectively)
and one simultaneous first order transition for $\epsilon \gg \epsilon_0$ (shown as a solid line). Near the region $\epsilon\simeq \epsilon_0$
(segment AB in the figure), 
the two transition temperatures can be very close and can be continuous or first order.
In iron pnictides, $\epsilon$ would refer to the larger of the spin anisotropies or 3D couplings.
}
\label{fig:sketch_phase_diagram}
\end{figure}

Based on the above, we propose a phenomenological
phase diagram (See Fig. \ref{fig:sketch_phase_diagram} ) with the following features:

1. The structural transition is driven by orbital ordering, which
happens at temperature $T_O$. It is set by the exchange energy scale
in the problem.

2. Let $\epsilon_0$ be the energy scale below which the long-wavelength
fluctuations are suppressed. Then the ratio of Neel to orbital transition
can be parametrized as ($x=\epsilon/\epsilon_0$)
\begin{equation}
{T_N\over T_O}={2-x \over 1 + \ln{1/x}} \qquad for~ x\ll1,
\end{equation}
and,
\begin{equation}
T_N/T_O=1 \qquad for~ x\gg1.
\end{equation}
We have two continuous phase transitions for $x\ll1$ and one
simultaneous first order transition for $x\gg1$. In between,
the region $x\simeq 1$ can have a small stretch where the
two transitions are practically inseparable but remain
continuous (AB in Fig. \ref{fig:sketch_phase_diagram}). The two transitions merge
into a first order transition at the point B in Fig. \ref{fig:sketch_phase_diagram}.

We last note that in principle, doping can be the source of another kind of
additional fluctuation which significantly reduces both spin and
orbital transition temperatures from the mean-field values. 
This can also lead to the separation of 
structural and magnetic transition as observed in many families of iron pnictides.

\subsection{Discussion of materials}
We next discuss the relevance of this study to various experimental findings in the iron-pnictide materials.

As mentioned previously, the parent compounds of the 1111 family have two separate 
second order phase transitions, while in the 122 family the two transitions are 
closer to each other in temperature. In the undoped BaFe$_2$As$_2$, the two 
transitions are slightly separated, where the structural transition starts as 
second order and is followed by a simultaneous first order jump both in the 
lattice distortion and magnetic transition at a lower temperature. 
\cite{Birgeneau_Ba122, Kim_Ba122, Birgeneau_review} On the other hand, 
in CaFe$_2$As$_2$ and SrFe$_2$As$_2$ the structural and magnetic phase 
transitions happen together as a single first order transition.
\cite{Firstorder_Ca122,Firstorder_Sr122, Suchitra_Sr122}

The three behaviors reported in different iron-pnictide materials are all captured by our phase diagram of a coupled spin-orbital Hamiltonian. In particular, phase transitions in the 1111 family correspond to the case when $\epsilon$ is small and away from $\epsilon_0$, where the structural and magnetic phase transitions are separated and of second order. On the other hand, phase transitions of CaFe$_2$As$_2$ and SrFe$_2$As$_2$ correspond to the case when $\epsilon$ is much larger than $\epsilon_0$, where the two transitions occur as a single first order transition.
The region $\epsilon\simeq \epsilon_0$ is relevant to BaFe$_2$As$_2$. In this case, the structural and magnetic transition temperatures can be very close and there is a tendency for the magnetic transition to become first order. This indicates that these materials are close to the boundary between the
distinct regions.

One can further ask which interaction term controls $\epsilon$ in the iron-pnictide materials. In this study we have investigated the role of spin space anisotropies described by Eq. (\ref{eqn:anisotropy_term}) or Eq. (\ref{eqn:single_ion_term}). Their effects on the transition temperature are
in essence the same as exchange couplings in the third direction.\cite{Kivelson,Antropov, Kamiya} Phenomenologically, $\epsilon$ would refer to the larger of the terms in determining the phase transitions.
It is known that the 122 family is more disperse in the third direction than the 1111 family. In particular, in the 122 family spin-wave spectra from neutron scattering are usually fitted with an additional 3D exchange coupling $J_c$, while for the 1111 materials $J_c$ is essentially zero. In BaFe$_2$As$_2$, the third-direction coupling is non-zero but also appears small; the reported $J_c/J_1$ is roughly $1\%$.\cite{neutron_Ba122_Matan, neutron_Ba122_Christianson}
Since a 2D Ising universality has been found for this material,\cite{wilson10}
a uniaxial spin anisotropy could be more important.
On the other hand, in CaFe$_2$As$_2$ and SrFe$_2$As$_2$, $J_c$ is more substantial and $J_c/J_1$ is estimated to be 10$\%$.\cite{neutron_Ca122, neutron_anisotropy_Sr122} Therefore, in these materials spin exchange coupling in the third direction could be the controlling factor for $\epsilon$.
 
We note that coupling to other degrees of freedom such as the lattice variable could also turn the isolated orbital or magnetic transition into first order.
However, besides the phase diagram, the orbital variables have proven indispensable in describing various other properties of iron-based superconductors such as the emergent transport anisotropies.\cite{transport_Chu_magnetic, transport_Chu_mechanical, transport_Tanatar, transport_Ying, transport_Kuo, CCC_resistivity, Lv_resistivity}
 
A modest orbital polarization has been reported by angle-resolved photoemission (ARPES) experiments
performed on the 122 family of iron pnictide materials.\cite{ARPES_Shimojima, ARPES_Lee, ARPES_Wang, ARPES_Yi}
This observation is crucial in explaining the striking phenomenon that in these materials the resistivity is smaller in the longer AFM axis.
\cite{transport_Chu_magnetic, transport_Chu_mechanical, transport_Tanatar, transport_Ying, transport_Kuo}
This unexpected behavior is striking especially because optical measurements indicate a smaller scattering rate along the shortened FM direction.\cite{optical_Dusza, optical_Lucarelli} It is the presence of an anisotropic effective mass due to a preferred occupation of $d_{xz}$ over $d_{yz}$ orbitals on the Fermi level that renders a better conducting pathway along the AFM direction.\cite{CCC_resistivity, Lv_resistivity}

As mentioned previously, with a preferential occupation of $d_{xz}$ orbitals over $d_{yz}$ orbitals, 
relativistic spin-orbit coupling can induce an orbital angular momentum in the xy plane and lead to a single ion anisotropy.
An excess population of $d_{xz}$ orbitals (through an induced $d_{xz}+i d_{xy}$ piece) can favor spins to point along
the x-axis while excess population of $d_{yz}$ orbitals can favor spins to
point along the y direction. We propose that this mechanism
is the reason why the observed directions of ordered spin moments are tied to antiferromagnetism and end up point along the AFM direction.

We last note that a possible orbital ordering has also been proposed for Fe$_{1+y}$Te$_x$Se$_{1-x}$ (the so-called 11 family of iron chalcogenides). \cite{discovery_11, Ashvin_OO}. In these materials, the ordered moments form a $(\pi/2,\pi/2)$ diagonal double stripe pattern, and the spin orientation points  toward the \emph{FM direction}.\cite{Neutron_11_Bao, Neutron_11_Li} Based on our discussion above, we believe this implies on the Fermi level a preferred population of Wannier functions whose orbital lobes point along the same direction. One direct consequence of this prediction is that \emph{in 11 iron chalcogenides resistivity is smaller in the FM direction}.\cite{Ashvin_OO, TKLee} This is indeed consistent with recent resistivity measurements.\cite{JH_11} The above prediction could be further tested by future ARPES and optical experiments on de-twinned iron chalcogenides.

\section{Conclusion}
In summary, we have studied finite-temperature phase transitions in a Hamiltonian of coupled Heisenberg spin and Ising orbital degrees of freedom. Using mean-field theory, Monte Carlo simulations, and general arguments we established the phase diagram of such a spin-orbital model and discussed its relevance to the iron-pnictide superconductors. We found that if spin rotational invariance is preserved, the magnetic transition temperature is pushed to zero in accord with the Mermin-Wagner theorem. In this case, there is only one single finite-temperature orbital phase transition which belongs to the 2D Ising universality class. By introducing spin space anisotropies into the Hamiltonian, spins can order at finite temperatures and the magnetic and orbital transitions are found to couple together and become first order. This phase diagram captures several observed behaviors in the 1111 and 122 families of iron pnictides. We also studied the case when relativistic spin-orbit coupling leads to a uniaxial anisotropy and found that the preferred spin-orientation is driven by orbital order. This explains why the direction of ordered moment in these materials is tied to their antiferromagnetism.

In the field of iron-based superconductors, there are several open questions that remain to be answered. It is interesting to further explore other experimental implications of model Hamiltonians with coupled spin and orbital degrees of freedom. For example, can fluctuations in orbital and/or spin variables account for various anomalous phenomena that occur above the structural and magnetic transition temperatures?\cite{L_Greene} What are the effects of orbital order and orbital fluctuations on twin boundaries? Are they related to the enhanced superconductivity at domain walls in these materials? \cite{domainSC_moler, domainSC_Xiao} Calculations to address these interesting open questions are areas of future study.

 \section*{acknowledgments}
The authors acknowledge discussions with N. Curro, J.-H. Chu, I. R. Fihser,
 R. Thomale, A. Diougardi, W. Pickett, R. M. Fernandes, and Y. Kamiya. 
R.A. and R.R.P.S. are supported by NSF Grant No. DMR-1004231. C.C.C. and T.P.D. are supported by the U.S. DOE under Contract No. DE-AC02-76SF00515.

\section{Appendix}

\subsection{Monte Carlo Simulations with Parallel Tempering}
We have used a parallel tempering Exchange Monte Carlo (EMC) method
to simulate our models. \cite{tempering,feedback} It is an efficient extended ensemble
simulation method that simulates multiple copies (replicas) of the system
simultaneously at different temperatures. Exchanges between replica
configurations are accepted or rejected in accordance with detailed balance. Replica exchange has been used to study systems spanning many fields
including strongly correlated systems, biological pathways, and
spin glasses. \cite{emc_review}
The advantage of these methods is that while at high
temperatures the system's memory is erased and when replicas go back to lower
temperatures they explore large phase space uncorrelated in monte carlo time.\cite{broad_histogram} 

\begin{figure}[t!]

\includegraphics[width=0.8\columnwidth,clip,angle=-90]{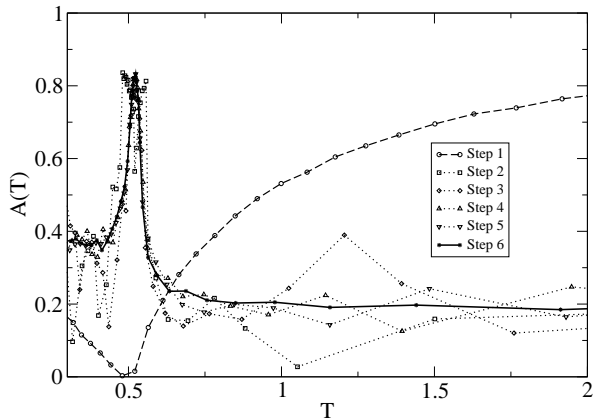}
\caption{The acceptance probability is plotted versus temperature for several feedback steps.
The initial geometric distribution shows a pronounced dip
in acceptance near the critical point and later feedback steps show how this is
corrected by clustering replicas around $T_c$.}
\label{plot}
\end{figure}

Recently, Katzgraber et al. \cite{feedback} showed that in order to maximally
benefit from EMC, the temperature distribution must be determined in a
nontrivial way via a "feedback" method. The temperature distribution $\{T_i\}$ is
obtained by starting with some initial set and recording statistics on the "round
trip" time from $T_{low}$ to $T_{high}$. Minimization of this round trip time
results in the optimal distribution.  
The endpoints,
$\{T_{low}...T_{high}\}$ are fixed and feedbacks of the simulation are done
until the distribution converges. The evolution of acceptance probability is
 shown in Fig.~\ref{plot}. Once $\{T_i\}$ is determined, an 
exchange monte carlo simulation is performed using the stored optimal 
temperature distribution.

Between the EMC moves that exchange replicas, one has some freedom in how to
update each individual replica, provided one can always know the energy of that
replica. We choose to do local spin/orbital flips by sweeping over the lattice
and randomly choosing whether or not a site in the lattice attempts
a spin flip or an orbital flip.

\subsection{Orbital and Spin Measurements}
There are two order parameters of interest for our Hamiltonian, one associated
with the orbital degrees of freedom and another with the spin degrees of
freedom. 
We measure second and fourth moments for both variables, and, in the case of
the magnetization, we measure at the two AFM wave vectors of interest,
$\vec Q_1 = (\pi,0)$, and $\vec Q_2 = (0,\pi)$.

\begin{align}\label{eqn:orbital_meas}
\langle n^2 \rangle = \langle \left( (\frac{1}{N}\displaystyle\sum_i^N n_i) -\langle n \rangle\right)^2\rangle\\
\langle n^4 \rangle = \langle \left( (\frac{1}{N}\displaystyle\sum_i^N n_i) -\langle n \rangle\right)^4\rangle
\end{align}

\noindent
\begin{align}\label{eqn:spin_meas}
\langle S^2 \rangle = \langle \left( \frac{1}{N} \displaystyle\sum_{i}^N \vec S_i e^{i\vec Q \cdot \vec r_i} \right)^2\rangle\\
\langle S^4 \rangle = \langle \left( \frac{1}{N} \displaystyle\sum_{i}^N \vec S_i e^{i\vec Q \cdot \vec r_i} \right)^4\rangle
\end{align}


\noindent
The $n_i$ take value 0 or 1 in our model and $\langle n\rangle = \frac{1}{2}$.
The $S_i$ are classical Heisenberg spins with magnitude unity and
$\langle\vec{S}\rangle = 0$. These orbital and spin measurements are used 
to evaluate Binder cumulant ratios as discussed below.

\subsection{Binder Ratios}
We define the Binder ratios for the
 orbitals $g_n$ and for the spins $g_S$ through the relations:

\begin{align}\label{eqn:binder}
g_n = \frac{\langle n^4 \rangle}{\langle n^2 \rangle^2}\\
g_S = \frac{\langle S^4 \rangle}{\langle S^2 \rangle^2}.
\end{align}

\noindent
At low temperatures, the spin and orbital order parameter distributions will be
sharply peaked at their extremum values. At high temperatures all variables will have gaussian
distributions.
\noindent
For $T \ll T_c$ the orbital quantities are:
\begin{align}
\langle n^2 \rangle = \frac{1}{4}\\
\langle n^4 \rangle = \frac{1}{16}\\
g_n = 1~.
\end{align}
\noindent
For the spins, the low temperature limits are 
\begin{align}
\langle S^2 \rangle = \frac{3}{2}\\
\langle S^4 \rangle = \frac{9}{2}\\
g_S = 2
\end{align}
\noindent
The two orbital orders divide the system between $Q_1$ and $Q_2$, 
resulting in different limits than a system without competing ordering wave vectors.

At high temperature, $T\gg T_c$, we get well known results
for the Binder ratio:
\begin{align}
\label{eqn:high_limits}
\langle n^2 \rangle = 1\\
\langle n^4 \rangle = 3\\
g_n = 3\\
\langle S^2 \rangle = 3\\
\langle S^4 \rangle = 15\\
g_S = \frac{5}{3}
\end{align}
\noindent
The difference between orbitals and spins comes purely
from the dimensionality of the variable.
We note that for $g_S$, in contrast to the low temperature limits, the high temperature
limits do not depend on the presence two ordering wavevectors.

\subsection{First and Second Order Phase Transitions}
We can estimate the thermodynamic $T_c$ by carefully studying the size dependence of
various physical quantities. We rely on the Binder ratios defined previously and
well known finite size scaling arguments to address the types of transitions we measure.
We propose the usual scaling ansatz for the susceptibility, 

\begin{align}
\label{eqn:scaling}
t = \frac{T-T_c}{T_c}\\
\chi(t,L)  = L^{\frac{\gamma}{\nu}} \tilde{\chi}(L^{\frac{1}{\nu}}|t|).
\end{align}

\noindent
$t$ is the reduced temperature, and $\chi(t,L)$
is the susceptibility per spin for a system of size $L$.
$\tilde{\chi}$ is some unknown but universal function and $\nu$ and $\gamma$ are
critical exponents which denote the power law divergence at $T_c$.

The Binder ratios $g_n$ and $g_S$ have the property that at $T_c$, they are independent
of system size, universal constants of the system. We find $T_c$ from Binder Ratio measurements for many pairs of systems of size $L$ and $2L$ and plotting versus temperature. A crossing for a given pair gives a constant and $T_c$. Size dependence of this constant exponentially decays versus the system size. \cite{finite_size_scaling} A more prominent size dependence occurs for the spins than orbitals in Fig.~\ref{fig:anisotropy_regress}. We extrapolate the size dependence to large $L$ by fitting the
 exponential decay. The y-intercept of this fit is the thermodynamic $T_c$.

\begin{figure}[t!]
\includegraphics[width=0.8\columnwidth,clip,angle=-90]{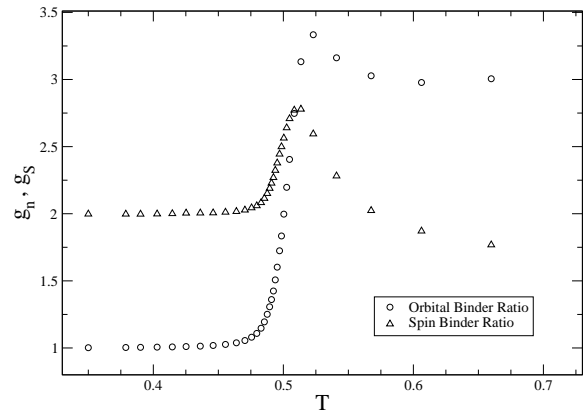}
\caption{For an anisotropic ( $\lambda = 0.75$ ) $24 \times 24$ system, we plot the orbital and spin binder ratios, $g_n$ and $g_S$. $g_n$ and $g_S$ both develop what seem like a divergence at the transition temperature. The development of such an incipient divergence is an indicator of a first order transition.}
\label{fig:binder_divergence_spinspace}
\end{figure}

Next we discuss the determination of the order of the transition that motivates
our phase diagram for the pnictides. At a second order phase transition, various thermodynamic quantities develop power law singularities characterized by
critical exponents, in this case $\nu$ and $\gamma$.
We arrive at Fig.~\ref{fig:critical_orb1} by varying critical exponents until a 
collapse of all points is achieved. In the case of anisotropy, there are no
critical exponents that produce a good data collapse, an indication that the transition is not second order. To support the claim that
the anisotropy leads to first order transitions, we show a plot of binder ratios
 for spin-space anisotropy ($\lambda = 0.75$) in Fig.\ref{fig:binder_divergence_spinspace}.
The binder ratio for spins develops a divergence near $T_c$ that is accompanied 
by a weak divergence for the orbital binder ratio. On its own, this method does
not conclusively establish  the first order nature of the
transition. However, in conjunction with the lack of critical exponents, we
propose the phase diagram in Fig. \ref{fig:sketch_phase_diagram}. Larger system sizes
would be helpful in further studying the divergence of binder ratios in Fig.\ref{fig:binder_divergence} and Fig \ref{fig:binder_divergence_spinspace}. \cite{1st_order_binder_div}


\end{document}